\documentclass[%
 reprint,
 amsmath,amssymb,
 aps,
pra,
]{revtex4-2}

\usepackage{graphicx}
\usepackage{dcolumn}
\usepackage{bm}

\begin{document}

\preprint{APS/123-QED}

\title{\bf Perfect soliton crystals on demand}

\author{Yang He}
\affiliation{Department of Electrical and Computer Engineering, University of Rochester, Rochester, NY 14627}
\author{Jingwei Ling}
\affiliation{Department of Electrical and Computer Engineering, University of Rochester, Rochester, NY 14627}

\author{Mingxiao Li}
\affiliation{Department of Electrical and Computer Engineering, University of Rochester, Rochester, NY 14627}

\author{Qiang Lin}
\email{qiang.lin@rochester.edu}
\affiliation{Department of Electrical and Computer Engineering, University of Rochester, Rochester, NY 14627}



\begin{abstract}
Recent advance of soliton microcombs has shown great promise to revolutionize many important areas such as optical communication, spectroscopic sensing, optical clock, and frequency synthesis. A largely tunable comb line spacing is crucial for the practical application of soliton microcombs, which unfortunately is challenging to realize for an on-chip monolithic microresonator. The recently discovered perfect soliton crystal (PSC) offers a convenient route to tune the comb line spacing. However, excitation of a PSC is generally stochastic by its nature and accessing a certain PSC state requires delicate tuning procedure. Here we demonstrate the on-demand generation of PSCs in a lithium niobate microresonator. The unique device characteristics allow us to produce a variety of PCSs and to switch between different PSC states, deterministically and repetitively. We utilize the device to show arbitrary dialing of the comb line spacing from 1 to 11 times of the free-spectral range of the resonator. The demonstration of PCSs on demand may now open up a great avenue for flexibly controlling the repetition rate of soliton pulses, which would significantly enhance and extend the application potential of soliton microcombs for communication, signal processing, and sensing.

\end{abstract}

\maketitle
\section{Introduction}

Optical frequency combs produced in monolithic microresonators, with compact device sizes and broad spanning spectra, have attracted remarkable attention in the past decade, which have been demonstrated on a variety of platforms  \cite{kippenberg2011microresonator, gaeta2019photonic}. One important class of Kerr frequency combs is soliton microcombs \cite{Kippenberg14, Vahala15, Maleki15, Gaeta162, Kippenberg162, Gaeta16, Wang2016, Diddams17, Kippenberg172, Kartik17, Tang18, He18}, in which all comb lines are locked in their phases, producing a coherent periodic pulse train in the time domain. The superior coherence property of soliton microcombs leads to many applications, such as communication \cite{Koos17}, spectroscopy \cite{Vahala162, Lipson18}, ranging \cite{Vahala18, Koos18}, and frequency metrology \cite{Newman2018}, with many others \cite{kippenbergreview2018} expected in the years to come. Unfortunately, the monolithic resonator platforms come with a significant cost in that the physical geometry of a resonator is defined during device fabrication, leading to a fixed comb line spacing (CLS) that is difficult to adjust once the device is made. This drawback limits the application potential of soliton microcombs since different applications have distinctive requirements on CLS. For example, optical communication requires a large CLS to fit wavelength-division multiplexed channels while spectroscopic sensing requires a small CLS to achieve high spectral resolution \cite{newbury2011searching}. A largely tunable CLS is thus indispensable for future applications of soliton microcombs. 

\begin{figure*}[t!]
	\centering\includegraphics[width=2\columnwidth]{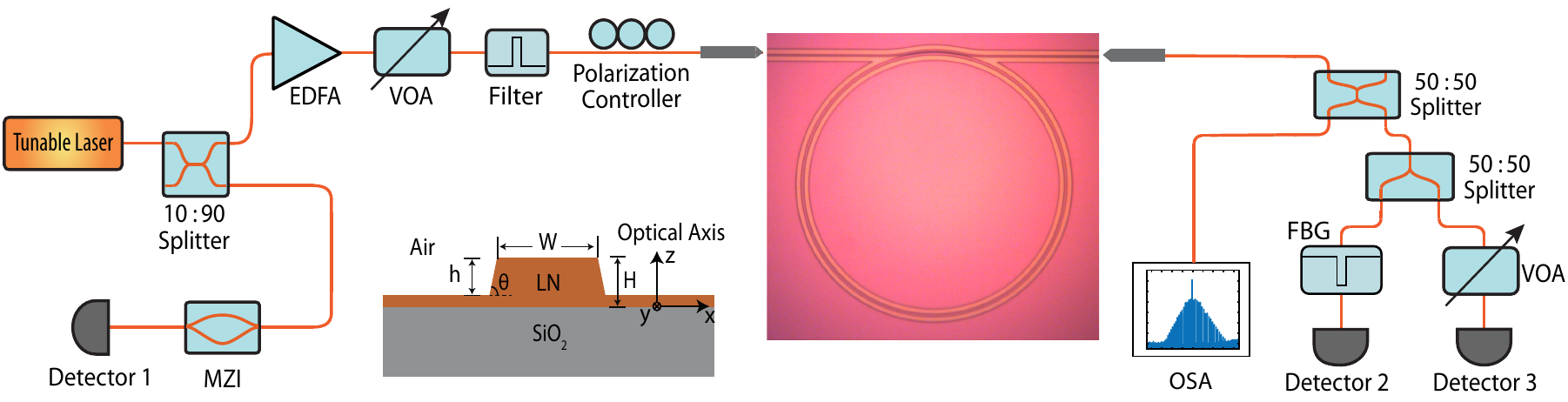}
	\caption{\label{Fig1} Schematic of the experimental setup, where the optical microscope image shows the LN microring resonator with a radius of 100~$\mu$m. The inset shows the schematic of the waveguide cross section of the LN microring, with W~=~2~${\mu m}$, H~=~600~nm, $~h=~410$~nm, and $\theta=75^\circ$. EDFA: erbium-doped fiber amplifier; VOA: variable optical attenuator;  MZI: Mach-Zehnder interfeometer; FBG: Fiber Bragg grating; OSA: optical spectrum analyzer. }
\end{figure*}

Recently, it was shown  \cite{Diddams17, wang18, Kippenberg19} that soliton microcomb exhibits intriguing crystallization processes where Kerr solitons are self-organized around the perimeter of a microresonator, forming various soliton-crystal states. Of particular interest is the perfect soliton crystal (PSC) free from defects \cite{Kippenberg19}, in which the underlying soliton pulses are evenly spaced in time within a round trip, with a CLS of $n \times {\rm FSR}$, effectively increased by a factor of an integer number $n$ from the free-spectra range (FSR) of the cold cavity. From the application point of view, the pulse train of a PSC is fundamentally identical to that of a single-soliton state produced in a smaller resonator. Therefore, PSCs offer a convenient approach for tuning the repetition rate of soliton micocomb in a monolithic resonator. However, excitation of a PSC state is often stochastic since the soliton formation is interfered by the significant thermo-optic nonlinearity \cite{Kippenberg14,Kippenberg17} and the stability regime of PSCs is surrounded by those of chaotic combs \cite{Kippenberg19}. Consequently, accessing and switching of PSCs requires a delicate tuning process in the two-dimensional space of pump power and frequency \cite{Kippenberg19}. This tuning process, moreover, can only change the soliton number of a PSC state in a decreasing order \cite{Kippenberg17, Kippenberg19}. So far, it remains a challenge in freely accessing PSCs with arbitrary values of $n$, which is crucial for the practical application of perfect soliton crystals.

Here we demonstrate the generation of PCSs \emph{on demand}, where the PCSs can be accessed and switched freely with desired values of CLS inside a single lithium niobate (LN) microresonator. This is enabled by the unique photorefractive property of the device, which was recently shown \cite{He18} to self-stabilize the soliton formation regime, allowing solitons to self-start and to be switched bi-directionally. In this paper, we show that such a unique device characteristics allows us to produce a variety of PCSs and to switch between different PSC states, deterministically and repetitively. As an example demonstration, we show continuous dialing of the comb line spacing from 1 to 11 times of the FSR of the resonator. The demonstration of PCSs on demand may now open up a great potential for flexibly tuning the repetition rate of soliton pulse train for diverse applications. 

\begin{figure*}[t!]
	\centering\includegraphics[width=1.80\columnwidth]{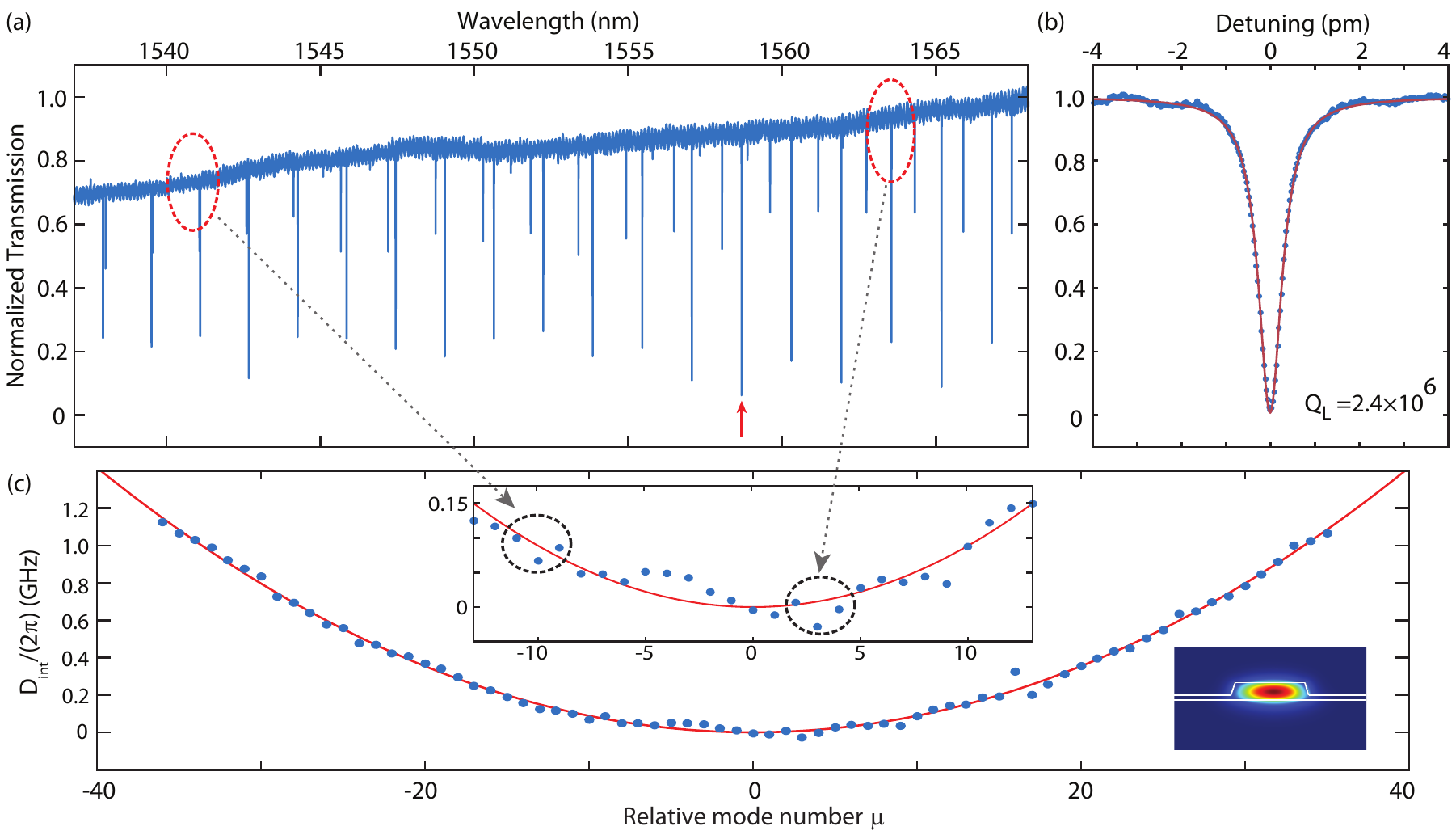}
	\caption{\label{Fig2} (a) Laser-scanned transmission spectrum of the LN microring. (b) The detailed transmission spectrum of the fundamental quasi-TE mode at 1558.68~nm, marked as red arrow in (a), with the experimental result shown in blue and the fitting result shown in red. (c) The recorded dispersion of the fundamental quasi-TE mode as a function of relative mode number ${\mu}$ (${\mu=0}$ is at 1558.68~nm). $D_{int} = \omega_\mu - \omega_0 - \mu D_1$ \cite{Vahala15}. The experimental data are shown as blue dots and the theoretical fitting curve is in red. The inset shows the zoom-in of the dispersion profile around ${\mu=0}$. }
\end{figure*}

\begin{figure*}[t!]
	\centering\includegraphics[width=1.9\columnwidth]{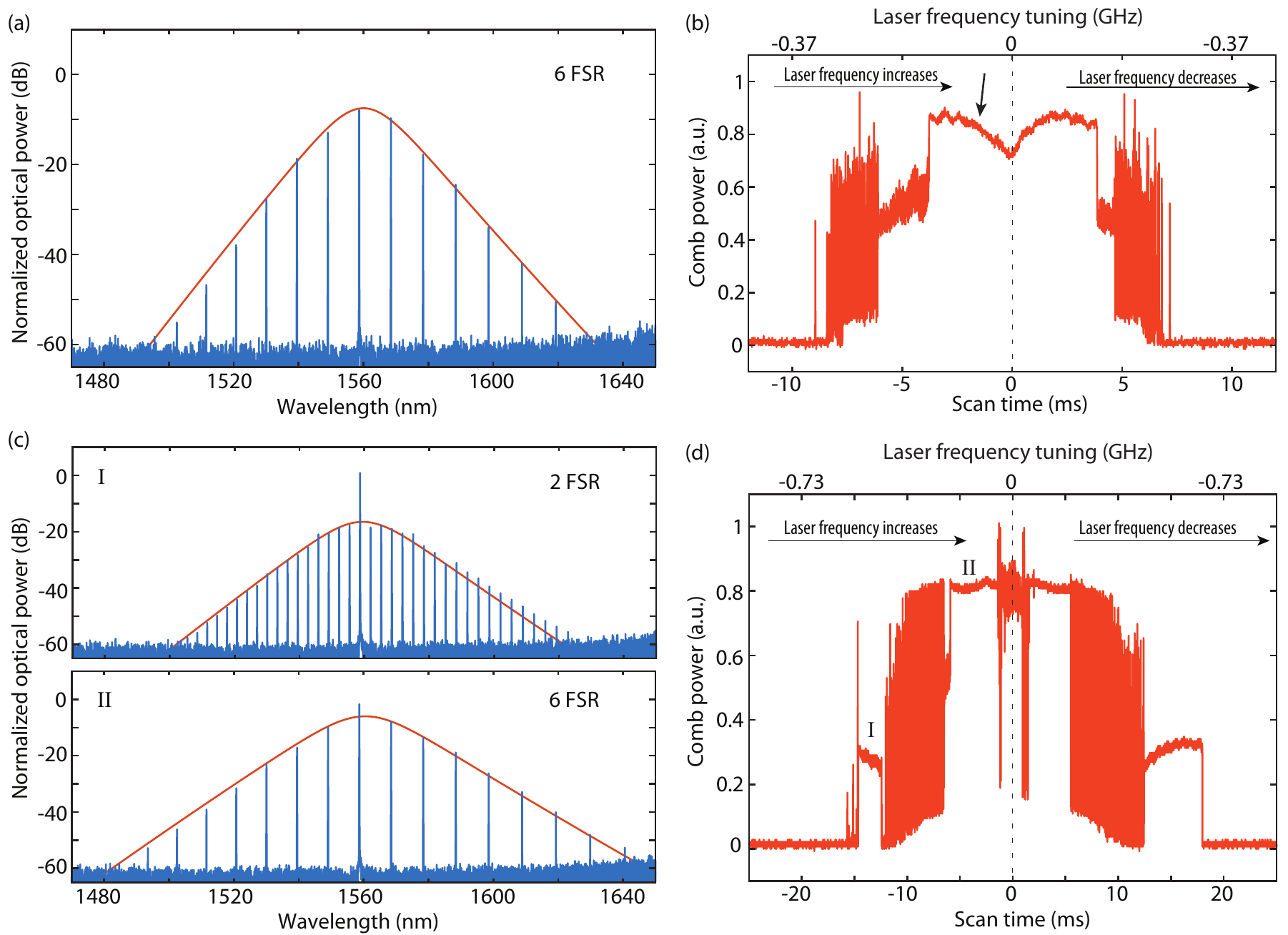}
	\caption{\label{Fig3} (a) Optical spectrum of a 6-soliton PSC state, recorded at the step indicated by the black arrow in (b). (b) Comb power steps when the laser frequency was scanned back and forth across the step (indicated by the black arrow) to produce the 6-soliton PCS shown in (a), with an on-chip pump power of 5.2~mW. (c) Optical spectrum of a 2-soliton (I) and 6-soliton (II) PCS states, recorded at the comb power steps I and II, respecitvely, indicated on (d). (d) Similar to (b), but with on-chip power of 8.3~mW.}
\end{figure*}

\section{Device properties}

The device employed for producing PSCs is a Z-cut LN microring resonator sitting on oxide, with a radius of 100~$\mu$m. An image of the device is shown in Fig.~\ref{Fig1} and the detailed device parameters are provided in the figure caption. To test the device properties and to produce the soliton combs, a telecom-band continuous-wave (CW) tunable laser was used as the pump, whose power was boosted by an erbium-doped fiber amplifier~(EDFA) (Fig.~\ref{Fig1}). After the amplifier noise was cut off with a tunable filter, the laser was launched onto the chip via a lensed fiber. The fiber-to-chip coupling loss is about 5~dB/facet. The device output was then collected by another lensed fiber whose output was split into three parts: 50\% was recorded by an optical spectrum analyzer (OSA) to characterize the comb spectrum. 25\% passed through a fiber Bragg grating (FBG) filter to remove the pump mode so as to record the sole comb power produced. Another 25\% was recorded by a detector to monitor the cavity transmission of the pump. The laser wavelength was calibrated with a Mach-Zehnder interferometer (MZI).

Figure \ref{Fig2}(a) shows a laser-scanned spectrum of the cavity transmission in the telecom band, where the fundamental quasi-TE modes of the resonator has a free-spectral range of $\sim$1.6~nm ($D_1/(2\pi) =$ 199.7~GHz) and exhibit a loaded optical Q $\sim$2.4 million as shown in Fig.~\ref{Fig2}(b). Detailed characterization of the dispersion for the fundamental quasi-TE mode (Fig.~\ref{Fig2}(c)) shows that it exhibits an anomalous dispersion of $D_2/(2\pi) = 1.76$~MHz, corresponding to a group-velocity dispersion of $\beta_2 = -0.047~{\rm ps^2/m}$. Figure \ref{Fig2}(c) shows that there exist various avoided mode crossings around this spectral band, with some examples shown in detail in the inset. For example, an avoided mode crossing appears at mode $\mu= -11$, which is induced by the mode interacation with the second-order quasi-TE mode as clearly visible in Fig.~\ref{Fig2}(a). Another one appears at mode $\mu= 3$ that is induced by the mode interaction with the third-order quasi-TE mode (Fig.~\ref{Fig2}(a)). These avoided mode crossings have important impacts on the generation of PSCs, which will be utilized in the following to engineer the properties of PSCs.

\section{Generation and switching of perfect soliton crystals}
To produce PSCs, we launched a pump power of 5.2~mW onto the chip and scanned the laser frequency back and forth at the red-detuned side of the cavity mode at 1558.68~nm. The photorefractive effect of the LN microresonator self-staibilizes the laser-cavity frequency detuning around this regime (see below), which allows us to freely access various soliton states. As shown in Fig.~\ref{Fig3}(b), When the pump frequency was tuned from red to blue, the produced frequency comb first exhibited dramatic fluctuations on its power and then transited into an intermediate state in which the power fluctuation was significantly suppressed. Finally, it settled to a stable state (indicated by an arrow in Fig.~\ref{Fig3}(b)), at which the comb spectrum shows a smooth ${\rm sech^2}$ envelope (Fig.~\ref{Fig3}(a), red curve), with a CLS of $6\times {\rm FSR}$. The comb spectrum exhibited a clean background, with a signal-to-noise ratio greater than 50~dB around the center of the spectrum. All these characteristics indicate that the stable comb corresponds to a 6-soliton PSC state. When the frequency of the pump laser was tuned backward from blue to red (Fig.~\ref{Fig3}(b)), the whole process was reversed, where the frequency comb transited from the stable PCS state to an intermediate state and then the noisy regime. The intermediate transition states are likely related to the onset of modulational instability and the passage through a soliton breathing stage \cite{Kippenberg19}. Their detailed dynamics, however, are too fast \cite{lucas2017breathing, yu2017breather} to be characterized by our current setup, which will be left for future exploration.

At this level of pump power, the comb always arrives at the same PCS state and the single-soliton state does not show up, which is consistent with a recent observation \cite{Kippenberg19}. This phenomenon can be understood qualitatively by a simple picture as follows. At a low pump power, only at a small laser-cavity detuning is the intracavity pump power large enough to support soliton formation. However, the small laser-cavity detuning enhances the beating between the soliton pulse tail and a dispersive wave induced by a certain avoided mode crossing \cite{parra2017interaction, wang2017universal}. Consequently, the significant temporal oscillation along the soliton tails initiates the production of other solitons and the locking of their temporal positions \cite{parra2017interaction, wang2017universal}, which in turn precludes the existence of the single-soliton state. If the oscillatory tail extends to a considerable temporal length (say, comparable to the round-trip time of the cavity), the temporally locked solitons naturally forms a deterministic PSC state. This simple picture also explains that it is easier to produce a PSC state with a large soliton number at a smaller laser-cavity detuning, as we will show in the following.

When the on-chip pump power was increased to 8.3~mW, the PSC can now be switched directly between two states. As shown in Fig.~\ref{Fig3}(d), when the laser frequency was tuned into the cavity resonance from the red-detuned side, the pump laser excited first a 2-soliton PSC state (Fig.~\ref{Fig3}(c)I). The PSC is likely related to the mode crossing with the third-order quasi-TE mode, as indicated by a circle around ${\mu=2}$ in the inset of Fig.~\ref{Fig2}(c). This PSC state then transited, again via fluctuating intermediate comb states, into a 6-soliton PSC state (Fig.~\ref{Fig3}(c)II) which is similar to that shown in Fig.~\ref{Fig3}(a) while with a broader spectral bandwidth. Again, both PSC states exhibit nice ${\rm sech^2}$ spectral envelopes and clean backgrounds free from any residual defect modes, indicating the high purity of produced PSC states. 

It was shown recently \cite{Kippenberg19} that the PSC states can be switched via a certain complex route in the two-dimensional space of pump power and frequency. Here, in contrast, a simple frequency tuning along the red-detuned side enables the direct PCS switching. In particular, this switching is bi-directional, meaning that the two PSC states can be switched back and forth repetitively with 100\% probability by tuning the pump frequency, as shown clearly in Fig.~\ref{Fig3}(d). The phenomena observed here reveals that our PSC states feature deterministic switching on demand.  

\begin{figure}[t!]
	\centering\includegraphics[width=1\columnwidth]{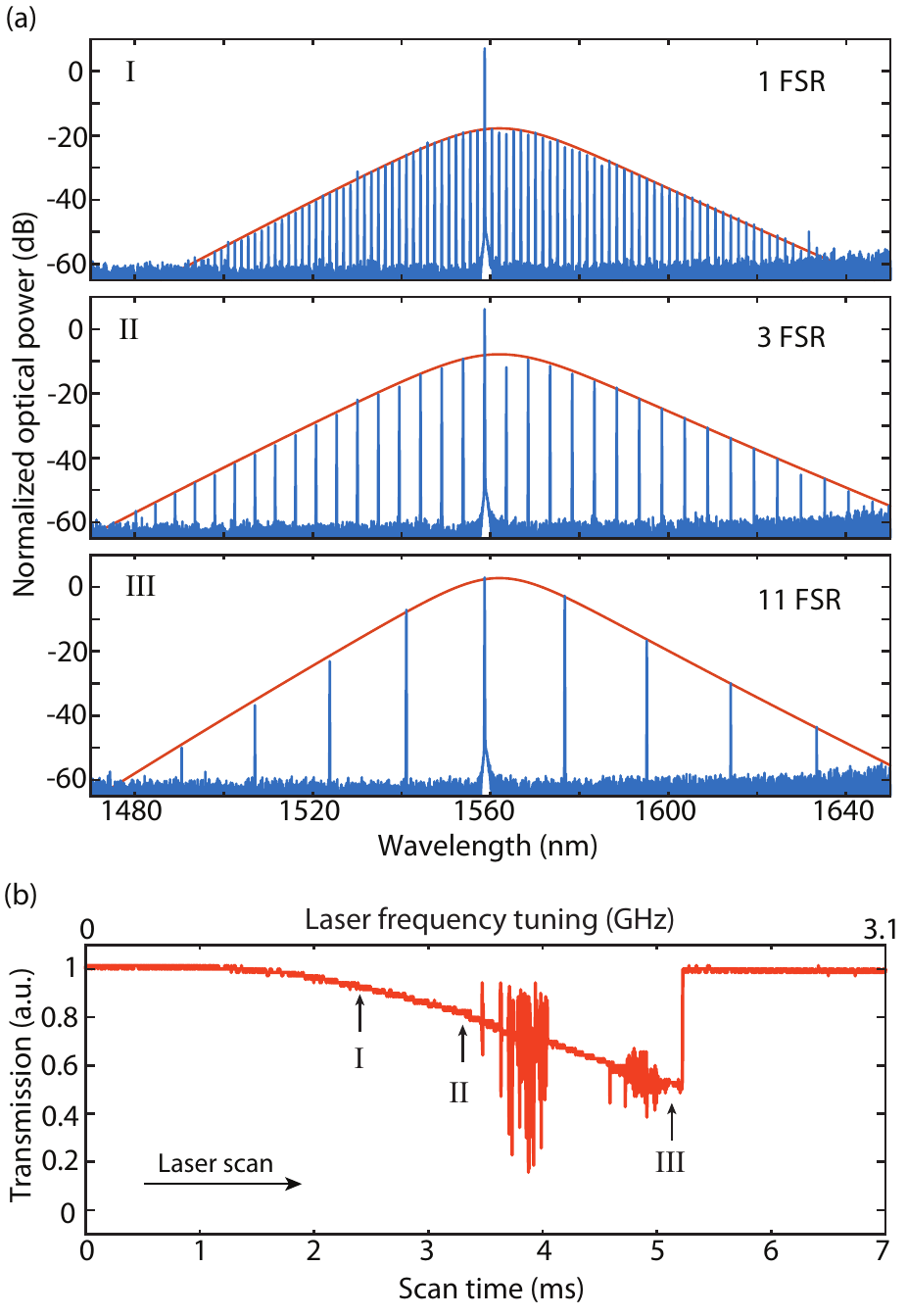}
	\caption{\label{Fig4} (a) Optical spectra of different PSC states, recorded at three laser-cavity detuning points indicated by arrows I, II, and III, respectively, in (b). (b) Laser-scanned transmission spectrum of the pump mode at 1558.68~nm (indicated by the red arrow in Fig.~\ref{Fig2}(a)), with an on-chip pump power of 12.4~mW.}
\end{figure}

\begin{figure*}[t!]
	\centering\includegraphics[width=2.0\columnwidth]{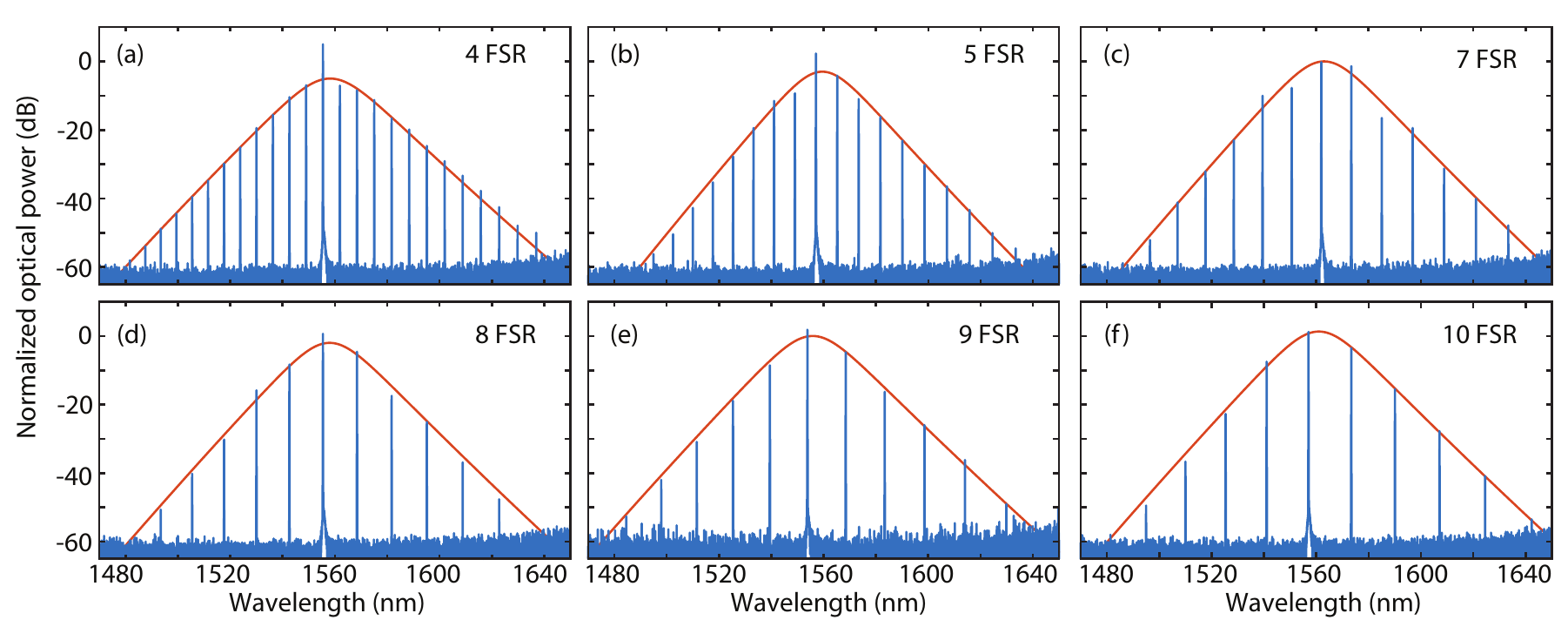}
	\caption{\label{Fig5} Optical spectra of PSC states with six different soliton numbers, recorded in the same LN microring resonator. }
\end{figure*}

Figure \ref{Fig3}(d) shows that the 6-soliton PSC state has a total comb power about 2.5 times of the 2-soliton PSC state. This is primarily because the increased number of solitons per round trip, as individual solitons in the two states exhibit similar spectral bandwidths (Fig.~\ref{Fig3}(c)I and II) and thus similar pulse energies \cite{coen2013universal}. Clearly, PSC with a large soliton number offers a convenient route for significantly increasing the power efficiency of soliton comb generation. 

The device enables even more versatile PSC switching at higher pump powers. Figure \ref{Fig4}(a) shows an example with a pump power of 12.4~mW on chip. To clearly show the laser-cavity frequency detuning of the pump wave, we plot in Fig.~\ref{Fig4}(b) the transmission trace of the pump mode. It exhibits a triangular shape facing towards high frequency, a result of the photorefractive effect in the z-cut LN microresonator. With the self-stabilized red laser-cavity detuning, we can now freely access and translate diverse PSC states, simply by tuning the pump frequency. For example, tuning the pump frequency to Point I indicated in Fig.~\ref{Fig4}(b) produces a single-soliton state as shown in Fig.~\ref{Fig4}(a)I. The appearance of the single-soliton state is because the large pump power supports the soliton formation at a far red detuned frequency. A large laser-cavity detuning, however, significantly damps the temporal oscillation along the soliton pulse tails \cite{parra2017interaction, wang2017universal}, which suppresses the probability of initiating other solitons and weakens the long-range temporal interaction, leading to high probability of producing a single-soliton state. When the laser-cavity detuning is decreased, the single-soltion state translates into a 3-soliton PSC state at Point II and then to a 11-soliton PSC state at Point III with a small laser-cavity detuning. The last state is likely related to the mode interaction around ${\mu=-11}$, as indicated by the circle in the inset of Fig.~\ref{Fig1}(c). A side benefit of producing PSC at a small laser-cavity detuning is that it helps remove the CW pump background along with the soliton comb, since the majority of the pump power is consumed inside the resonator. This is clearly evident by the negligible extra pump power in the 11-soliton state shown in Fig.~\ref{Fig4}(a)III compared to that in the single-soliton state shown in Fig.~\ref{Fig4}(a)I. \if{The switching between multiple PSC states is quite universal and can be flexibly accessed with different pump modes. As shown in Fig.~\ref{Fig4}(c) and (d), very similar phenomenon was obtained by switching to pump at a nearby mode at 1557.05nm, where the PSCs can be translated between 1-, 3-. and 10-soliton PSC states.}\fi 

The deterministic accessing and versatile switching of diverse PSC states in the device enables obtaining PSC states \emph{on demand} with a large range of soliton numbers inside a single device. To demonstrate this capability, we show in Fig.~\ref{Fig5} different PSCs with diverse CLSs. By simply choosing appropriate power and laser-cavity detuning, as well as cavity mode of the pump wave, PSC states with 4-, 5-, 7-, 8-, 9-, and 10-solitons are obtained. As soon as these three parameters are determined for a certain PSC state, the state can be accessed deterministically on demand every time by simply dialling in these parameters to the device. Together with those shown in Fig.~\ref{Fig3} and \ref{Fig4}, the soliton number of the PSC state can be continuously dialed from 1 to 11, clearly showing the power of our device for producing PSCs.

\section{Conclusion and discussion}

In summary, we have demonstrated the on-demand generation of a variety of perfect soliton-crystal states in a single z-cut LN microresonator. The unique device characteristics not only enables free access to desired PSC states deterministically and repetitively, but also allows to switch and translate between different PSC states bi-directionally. With the single device, we have shown diverse PSC states whose soliton number can be tuned arbitrarily from 1 to 11. These demonstrations clearly show the powerful capability of LN microresonators for on-demand generation of PCSs, which would significantly enhance and extend the application potential of soliton microcombs. For example, we can expect that a single LN microresonator can be easily reconfigured to produce versatile PSCs to fit the different demands of optical communications in either short-reach data-communication that requires only coarse wavelength division multiplexing (WDM) or in long-haul coherent communication where dense WDM is desired \cite{agrawal2012fiber}. The tuning resolution of the CLS for PSCs is fundamentally determined by the FSR of the microresonator. Our current device has a FSR of 199.7~GHz which could be too large for some applications. We expect to develop in the near future large LN microresonators with FSR in the range of 1-10GHz \cite{suhastrocomb2018} that would fulfill most of application needs in practice. On the other hand, lithium niobate exhibits strong electro-optic effect which has recently been utilized for a variety of electro-optic functionalities \cite{chen2014hybrid, chiles2014mid, rueda2016efficient, weigel2018bonded, wang2018integrated, rueda2019resonant, zhang2019broadband, zhang2019electronically, wang2019monolithic}. Integration of electro-optic modulation with the tunable PSCs on a monolithic LN platform would open up a great potential for broad applications in communication, signal processing, and sensing, that are expected to be realized in the near future.

\section*{Funding Information}
Defense Threat Reduction Agency-Joint Science and Technology Office for Chemical and Biological Defense (grant No.~HDTRA11810047); National Science Foundation (NSF) (EFMA-1641099, ECCS-1810169, and ECCS-1842691).

\section*{Acknowledgments}

This work was performed in part at the Cornell NanoScale Facility, a member of the National Nanotechnology Coordinated Infrastructure (National Science Foundation, ECCS-1542081), and at the Cornell Center for Materials Research (National Science Foundation, DMR-1719875). 
\bibliography{ref}

\end{document}